\documentclass[twocolumn,showpacs,preprintnumbers,amsmath,amssymb]{revtex4}
\usepackage{graphicx}

\begin{document}

\title{Searches on star graphs and equivalent oracle problems}
\author{Jaehak Lee$^1$, Hai-Woong Lee$^1$, and Mark Hillery$^2$}

\affiliation{
$^{1}$Department of Physics, Korea Advanced Institute of Science and Technology, Daejeon 305-701, Korea\\
$^{2}$Department of Physics, Hunter College of the City University of New York, 695 Park Avenue, New York, NY 10021 USA}

\begin{abstract}
We examine a search on a graph among a number of different kinds of objects (vertices) one of 
which we want to find.  In a standard graph search, all of the vertices are the same, except for one,
the marked vertex, and that is the one we wish to find.  We examine the case in which the unmarked
vertices can be of different types, so that the background against which the search is done is not
uniform.  We find that the search can still be successful, but the probability of success is lower than
in the uniform background case, and that probability decreases with the number of types of unmarked
vertices.  We also show how the graph searches can be rephrased as equivalent oracle problems.
\end{abstract}

\maketitle

\section{Introduction}
One of the main reasons quantum walks were originally introduced into 
quantum information was the hope that they might provide a 
useful conceptual framework for the development of new quantum algorithms \cite{aharanov}.  
This hope has been fulfilled \cite{ambainis07}-\cite{FaGoGu07}.  In addition, they have recently 
attracted considerable experimental interest.   They have been realized in optical lattices \cite{Dur2002,KaFoChStAlMeWi09} and with photons in waveguide lattices \cite{PeLaPoSoMoSi08}. 
Analogs of quantum walks, which are based on the wave nature of classical light, make use of the fact 
that the most important element in the behavior of a quantum walk is interference.  These were 
proposed in \cite{KnRoSi03,JePaKi04} and experimentally realized in Ref.~\cite{BoMaKaScWo99}. 
The use of interferometers to realize quantum walks was proposed \cite{JePaKi04} and recently experimentally implemented \cite{ScCaPo09}. Quantum walks with trapped ions have been theoretically described \cite{Travaglione2002,Xue2009} and experimentally realized as well 
\cite{ScMaScGlEnHuSc09,ZaKiGeSoBlRo09}. Finally, experiments in \cite{BrFeLaKaAsWh10}  and
\cite{Peruzzo} provide a framework in which walks with more than one walker can be studied. 

Here we will look at quantum walks on star graphs and equivalent oracle problems.  A star graph
has a single central vertex with $N$ edges emanating from it.  Each of these edges is connected
to a single external vertex.  Thus the graph resembles spokes extending from a hub.  We shall
denote the central vertex by $0$ and the external vertices by $1$ through $N$.  Star graphs form
a natural arena in which to study quantum searches.  Some of the external vertices may have 
different properties from the others, and our task is to find these vertices.  In a standard quantum
search one object has different properties than all of the others, all of which are the same, and which
constitue a kind of background.  Here we want to examine searches on more complicated
backgrounds.  Suppose that we want to find an object with a particular property, but the other objects
that we must search among are not all the same.  In the case of a star graph, this could be realized
by having the external vertices reflect the walking particle with one of three phases.  One vertex  
(the marked vertex) reflects with the phase we want to find, and each of the other vertices reflects 
with one of the other two phases.  As we shall see, it is possible to find the marked vertex using a 
quantum walks, but there is a limit to the probability of successfully doing so.

We shall employ the scattering model of a quantum walk in which the particle resides on the edges 
of the graph and at each time step scatters at the vertices \cite{hillery1,hillery2}.  
Each edge has two states, corresponding to the two
directions a particle on it can be going.  In the case of a star graph, the edge between vertex $0$ and
$j$ is associated with the state $|0,j\rangle$, which corresponds to the particle being on that edge
and going from $0$ to $j$, and $|j,0\rangle$, which corresponds to the particle being on that edge
but going from $j$ to $0$.  The states corresponding to the edges form an orthonormal basis for
the Hilbert space of the particle on the graph.  

In addition to a Hilbert space we need a unitary operator that advances the walk one time step.  We
shall give a detailed description of this operator shortly, however when a particle scatters at an
external vertex the effect of the operator is to reflect the particle with a phase, and the phases can
be different at different vertices.  It is these phases that determine the behavior of the walk.  One 
case is where all of the phases are $0$ except for one which is $\pi$.  This simply reproduces the
standard Grover search, and the particle becomes localized on the edge with the $\pi$ phase
shift after $O(\sqrt{N})$ steps.  As we mentioned, we can do a search on a more varied background
by allowing the phases to have more values.
If roughly half have phase shifts of $2\pi /3$, roughly another half have
phase shifts of $-2\pi /3$, and a few have phase shifts of zero, the particle will become localized on
the edges with phase shift $0$ after $O(\sqrt{N})$ steps, but the maximum probability of finding it
there is not close to one, as in the case of the Grover search, but $3/4$.  Adding further variety to 
the phases causes this probability to decrease.  

We will begin by examining the eigenvalues of the unitary operator that advances the walk one step.
One result is that if most of the phase shifts are the same and only a few are different, 
the eigenstates group themselves into two subspaces, a big, relatively  boring, subspace in
which not much happens, and a small space in which interesting things can happen.  This allows
one to focus on what is happening in the small space and to ignore the large one, which has the
effect of simplifying the analysis of the walk.

Next, in order to connect the quantum walk description of a search to the usual oracle-based one,
we analyze what happens when the function the oracle evaluates is not a binary-valued one.  We 
examine two different uses of the oracle.  The first way of using the oracle reduces the problem to
a standard Grover search, and the probability of finding the marked vertex is close to one.  This 
use of the oracle, however, is not equivalent to what happens in our quantum walk.  We, therefore, 
explore a second way of using the oracle, which is equivalent to the quantum walk on the star graph.
Using the oracle in this way, for a function that can take more than two values,  corresponds to allowing 
more than two phases in the quantum walk.  For this use of the oracle, we further study how increasing 
the number of possible function values affects the number of steps in the search and the maximum probability
of finding the marked element.  These results carry over immediately to the quantum walk search on the
star graph with multiple phases and give us the number of steps required to localize the particle and 
the extent to which it can become localized.

\section{Eigenstates and eigenvalues of star graphs}
Let us now specify the details of the walk on a star graph.  As noted in the introduction, if the
star graph has $N$ edges,  we denote the central vertex by $0$ and the other vertices by
$1$ through $N$.  When the particle making the walk is reflected from the vertex at $j$, it picks up
a phase factor $e^{i\phi_{j}}$.  The action of the unitary operator that advances the walk one step
is given by
\begin{align}
U|0,j\rangle & = e^{i\phi_{j}} |j,0\rangle ,  \nonumber \\
U|j,0\rangle & = -r |0,j\rangle + t\sum_{k=1, k\neq j}^{N} |0,k\rangle \nonumber \\
& = -|0,j\rangle + t \sum_{k=1}^{N}|0,k\rangle ,
\end{align}
where $r=(N-2)/N$ and $t=2/N$.  Let
\begin{equation}
|\psi\rangle = \sum_{j=1}^{N} (\alpha_{j}|j,0\rangle + \beta_{j}|0,j\rangle ) .
\end{equation}
If $|\psi\rangle$ is an eigenvector of $U$ with eigenvalue $e^{i\theta}$, i.e.\ $U|\psi\rangle = 
e^{i\theta}|\psi\rangle$, then we have that
\begin{align}
-\alpha_{j} + t \sum_{k=1}^{N} \alpha_{j} & = e^{i\theta}\beta_{j} \nonumber \\
e^{i\phi_{j}}\beta_{j} & = e^{i\theta} \alpha_{j}  .
\end{align}
Solving the second equation for $\beta_{j}$ and inserting it into the first gives
\begin{equation}
-\alpha_{j} + tS = e^{i(2\theta -\phi_{j})}\alpha_{j}  ,
\end{equation}
where $S=\sum_{j=1}^{N} \alpha_{j}$.  There are now two cases.  First, suppose $S\neq 0$.  Then
solving the above equation for $\alpha_{j}$ and summing it over $j$ gives
\begin{equation}
S=tS \sum_{j=1}^{N} \frac{1}{(e^{i(2\theta -\phi_{j})} +1) }  ,
\end{equation}
so that the eigenvalues are determined by the equation 
\begin{equation}
\label{eigen1}
1= t \sum_{j=1}^{N} \frac{1}{(e^{i(2\theta -\phi_{j})} +1) } .
\end{equation}
If $S=0$ we have that 
\begin{equation}
-\alpha_{j} = e^{i(2\theta -\phi_{j})} \alpha_{j} ,
\end{equation}
which implies, if $\alpha_{j }\neq 0$, that 
\begin{equation}
\label{eigen2}
e^{i(2\theta -\phi_{j})} = -1 .  
\end{equation}
The only way this can be satisfied is if two or more
of the $\phi_{j}$'s have the same value.

Let's look at an example.  Suppose $\phi_{1} = \pi$ and $\phi_{j}=0$ for $j\geq 2$.  This is, of course,
the standard Grover case.  Setting $z=e^{2i\theta}$ Eq.\ (\ref{eigen1}) becomes
\begin{equation}
z^{2}-2rz+1=0 ,
\end{equation}
giving $z=r \pm i(1-r^{2})^{1/2}$.  Setting $e^{2i\theta_{0}} = r + i(1-r^{2})^{1/2}$, we find that 
this gives a value for $\theta_{0}$ that is of order $1/\sqrt{N}$.  We thus
have four eigenvalues of this type, $\pm e^{i\theta_{0}}$ and $\pm e^{-i\theta_{0}}$.  Eq.\ (\ref{eigen2}) 
yields the eigenvalues $\pm i$.  Each of these eigenvalues is $(N-2)$-fold degenerate.  It is also quite
straightforward to find the eigenvectors.  For $e^{i\theta_{0}}$ we find
\begin{align}
|\xi_{1}\rangle = {} & \eta \left[ \frac{1}{1-e^{2i\theta_{0}}} \left(|1,0\rangle - e^{i\theta_{0}}|0,1\rangle \right) \right. \nonumber \\
& \left. + \sum_{j=2}^{N} \frac{1}{1+e^{2i\theta_{0}}} \left(|j,0\rangle + e^{i\theta_{0}}|0,j\rangle \right) \right] ,
\end{align}
and for $-e^{i\theta_{0}}$
\begin{align}
|\xi_{2}\rangle = {} & \eta \left[ \frac{1}{1-e^{2i\theta_{0}}} \left(|1,0\rangle + e^{i\theta_{0}}|0,1\rangle \right) \right. \nonumber \\
& \left. + \sum_{j=2}^{N} \frac{1}{1+e^{2i\theta_{0}}} (|j,0\rangle - e^{i\theta_{0}}|0,j\rangle )\right] .
\end{align}
The expression for the eigenvector $|\xi_{3}\rangle$, which corresponds to eigenvalue 
$e^{-i\theta_{0}}$, is the same as that for $|\xi_{1}\rangle$ except with $\theta_{0}$ replaced by
$-\theta_{0}$,  and the expression for $|\xi_{4}\rangle$, which corresponds to eigenvalue 
$-e^{-i\theta_{0}}$, is the same as that for $|\xi_{2}\rangle$ except with $\theta_{0}$ replaced by
$-\theta_{0}$. 
The normalization constant $\eta$ is given by
\begin{equation}
\eta = \frac{\sin (2\theta_{0})}{ [N-(N-2)\cos (2\theta_{0}) ] ^{1/2}} \cong \frac{1}{\sqrt{N}} .
\end{equation}
The eigenvectors corresponding to the eigenvalues $\pm i$ are just determined by the condition 
$\sum_{j=2}^{N}\alpha_{j}=0$, so there is a great deal of freedom in choosing them.  One
possible choice for the eigenvectors corresponding to $+i$ is
\begin{equation}
|\zeta^{(+)}_{m}\rangle = \frac{1}{\sqrt{2(N-1)}} \sum_{j=2}^{N} e^{2\pi i(j-1)m/(N-1)}(|j,0\rangle
+i |0,j\rangle ) ,
\end{equation}
for $m=1,2,\cdots N-2$. The eigenvectors corresponding to $-i$ can be taken to be of similar
form.  If we start the walk in the state 
\begin{equation}
|\psi_{0}\rangle = \frac{1}{\sqrt{2N}} \sum_{j=1}^{N} (|j,0\rangle + |0,j\rangle ) ,
\end{equation}
we find that $|\psi_{0}\rangle \cong (|\xi_{1}\rangle + |\xi_{3}\rangle )/\sqrt{2}$, assuming $N\gg 1$,
so that the whole walk takes place in the space spanned by $|\xi_{1}\rangle$ and $|\xi_{3}\rangle$. 
After $m$ steps the state of the particle is
\begin{equation}
|\psi_{m}\rangle = \frac{1}{\sqrt{2}}(e^{im\theta_{0}}|\xi_{1}\rangle + e^{-im\theta_{0}} |\xi_{3}\rangle )
+O(N^{-1/2}) ,
\end{equation}
and when $m\theta_{0}=\pi /2$, we have  
\begin{equation}
|\psi_{m}\rangle =-\frac{1}{\sqrt{2}} (|1,0\rangle -|0,1\rangle ) +O(N^{-1/2}) .
\end{equation}
Therefore, to good approximation, the particle is located on the edge connected to the vertex with 
the $\pi$ phase shift, just as in the Grover algorithm.

Now let us move on to a more complicated situation.  Suppose that each of the $\phi_{j}$'s takes
on one of three values $\phi_{1}$, $\phi_{2}$ or $\phi_{3}$, and, in particular, $n_{1}$ are $\phi_{1}$,
$n_{2}$ are $\phi_{2}$, and $n_{3}$ are $\phi_{3}$, where $n_{1}+n_{2}+n_{3} = N$.  Then
Eq.\ (\ref{eigen1}) gives us
\begin{equation}
\frac{N}{2} = \sum_{j=1}^{3}\frac{n_{j}}{ze^{-i\phi_{j}}+1} ,
\end{equation}
which becomes
\begin{widetext}
\begin{equation}
\label{cubic}
z^{3} -2\left[ \sum_{j=1}^{3}x_{j}e^{i\phi_{j}} \right] z^{2} + 2\left[ x_{1}e^{i(\phi_{2}+\phi_{3})} + x_{2}e^{i(\phi_{1}+\phi_{3})} + x_{3} e^{i(\phi_{1}+\phi_{2})} \right]z - e^{i(\phi_{1}+\phi_{2}+\phi_{3})} =0 ,
\end{equation}
\end{widetext}
where $x_{j} = n_{j}/N$.  

One case in which the equation can be solved is if  $\phi_{1}=2\pi /3$, $\phi_{2}=-2\pi /3$ and 
$\phi_{3}=0$.  In addition, we shall assume that $N$ is odd, and $n_{1}=n_{2}= (N-1)/2$, and
$n_{3}=1$.  In that case, the above equation becomes
\begin{equation}
z^{3} + \left(\frac{N-3}{N}\right) z^{2} -  \left(\frac{N-3}{N}\right) z -1 =0 .
\end{equation}
The solutions to this equation are $z=1$ and
\begin{equation}
z=-1 + \frac{3}{2N} \pm i \left( \frac{3}{N} - \frac{9}{4N^{2}}\right)^{1/2}  .
\end{equation}
Note that the last pair of solutions for $z$ yields eigenvalues of the form $\pm [ i \pm O(N^{-1/2}) ]$.  
Comparing to the previous case we can surmise that this situation will again lead to a search in 
which we can find the vertex corresponding to the phase of zero in $O(\sqrt{N})$ steps.  This is, in
fact the case, but rather than use the exact eigenvalues to show this, we shall make use of an
approximation method which allows us to find a solution in a more general case.

We want to look at the situation in which one of the $x_{j}$'s is
small, typically of order $1/N$, and $N\gg 1$.    In that case the above equation can be solved
perturbatively.  If the full third degree polynomial above is $f(z)$, then let $f_{0}(z)$ be the same
polynomial but with the terms of order $1/N$ set equal to zero, that is, $f_{0}(z)$ is the $N\rightarrow
\infty$ limit of $f(z)$.  We then get a zeroth order solution to $f(z)=0$, which we shall denote by
$z_{0}$, by solving $f_{0}(z)=0$.

We can then get corrections to the zeroth order solution, by setting $z=z_{0}+\delta z$ 
substituting back into the cubic equation, keeping the lowest order surviving
terms, and then solving for $\delta z$.  This will give us
\begin{align}
\label{approx}
0 = {} & f(z_{0}+\delta z) \nonumber \\
\cong {} &  f(z_{0}) + [ f^{\prime}_{0}(z_{0}) + O(1/N) ]\delta z \nonumber \\
& + [ f^{\prime\prime}_{0}(z_{0})+   O(1/N) ] (\delta z)^{2} + \cdots .
\end{align}
Note that we have substituted $f_{0}+O(1/N)$ for $f$ in the derivative terms, 
e.g.\ $f^{\prime}_{0}(z_{0})  + O(1/N)$ for $f^{\prime}(z_{0})$, 
because the $f_{0}$ terms are typically of order one.  Because  $f(z_{0})$ is usually of 
order $1/N$ and the $(\delta z)^{2}$ term can be neglected in comparison to the $\delta z$ term, we
will usually have $\delta z \sim 1/N$.  This situation is not interesting, and let us see why.  
If $z_{0}=e^{i\theta_{0}}$, and $z_{0}+\delta z = e^{i(\theta_{0}+\delta\theta )}$, then $\delta z \sim 1/N$
will give us $\delta \theta \sim 1/N$.  After $m$ steps we will have
a factor of $e^{im(\theta_{0}+\delta\theta )}$ multiplying the eigenvector.  The part of this phase factor
that contains the information about the unusual vertices (the ones with the rare phase shift) is
$e^{im\delta\theta}$ and this phase factor will differ substantially from
one when $m$ is of order $N$.  That means that in this case, the unusual vertices would have a
significant effect on the dynamics after $N$ steps.
However, we could find the unusual vertices classically just by checking 
each vertex to see if its phase shift was one of the rare ones, and this procedure would take $N$
steps too, so we have gained nothing.

What we really want, to get some kind of a quadratic speedup, is for $\delta\theta \sim 1/\sqrt{N}$.  
Examining Eq.\ (\ref{approx}), we see that this can happen if the order one part of the 
coefficient of $\delta z$, which is $f^{\prime}_{0}(z_{0})$, vanishes.  We then get a quadratic
equation for $\delta z$
\begin{equation}
0 =  f(z_{0})+ [ O(1/N) ]\delta z +  f^{\prime\prime}_{0}(z_{0}) (\delta z)^{2} ,
\end{equation}
which typically results in  $\delta z$ being of order $N^{-1/2}$.  The condition 
$f^{\prime}_{0}(z_{0}) = 0$ implies that $z_{0}$ is a double root of the equation $f_{0}(z_{0})=0$.
Summarizing, in order to find a situation in which we can have a quadratic speed up, we want the
solution of the zeroth order problem to be a double root.

Let's give an example where that happens.  We are going to assume that almost half of the phases
are $\phi_{1}=2\pi /3$, almost half are $\phi_{2}=-2\pi /3$ and the remainder are $\phi_{3}=0$.  
We will take most of the phases from $j=1$ to $j=N/2$ to be $2\pi /3$, and most of the phases 
between $j=(N/2)+1$ to $j=N$ to be $-2\pi /3$, and we will split the phases that are zero 
putting some between  $j=1$ to $j=N/2$ and some between $j=(N/2)+1$ to $j=N$.
In particular, let us assume that for $j=1, \cdots, n_{31}$ and for $j=(N/2)+1, \cdots, (N/2)+n_{32}$
the phases are $0$, for $j=n_{31}+1, \cdots, (N/2)$ the phases are $2\pi /3$ and for $j=(N/2)+n_{32}+1,
\cdots, N$ the phases are $-2\pi /3$.  Therefore, the total number of edges with phase $0$ is
$n_{3}=n_{31}+n_{32}$, the total number with phase $2\pi /3$ is $n_{1}=(N/2)-n_{31}$ and
the total number with phase $-2\pi/3$ is $n_{2}=(N/2)-n_{32}$.  We will consider the case where
$n_{3} \ll N$.

Let us first get the zeroth order solution to Eq.\ (\ref{cubic}).  To do so, we neglect all quantities of
order $n_{3}/N$, which means we set $x_{1}=x_{2}=1/2$ and $x_{3}=0$.  Doing so, we get the
equation
\begin{equation}
z^{3}+z^{2}-z-1=(z^{2}-1)(z+1)=0 ,
\end{equation}
which has a double root at $z=-1$.  Let us now calculate corrections to this root.  First we set
$z=-1+\delta z$ and substitute it back into the equation.  We then keep terms of up to second order
in small quantities, with both $\delta z$ and anything of order  $n_{3}/N$ being considered small.
Toward this end, we set $x_{31}=n_{31}/N$ and $x_{32}=n_{32}/N$, and note that
\begin{equation}
 \sum_{j=1}^{3}x_{j}e^{i\phi_{j}} = -\frac{1}{2} + x_{31}(1-e^{2\pi i/3})+x_{32}(1-e^{-2\pi i/3}) ,
 \end{equation}
 and
 \begin{align}
 & x_{1}e^{i(\phi_{2}+\phi_{3})} + x_{2} e^{i(\phi_{1}+\phi_{3}} + x_{3} e^{i(\phi_{1}+\phi_{2})} 
 \nonumber \\
 & =  -\frac{1}{2} + x_{31}(1-e^{-2\pi i/3})+x_{32}(1-e^{2\pi i/3})  .
 \end{align}
Upon making our substitution and keeping only quantities of up to second order, we first find that,
as expected, the term proportional to $\delta z$ vanishes, and we are left with
\begin{align}
& (\delta z)^{2} - \{ x_{31}[ 2(1-e^{2\pi i/3}) + (1-e^{-2\pi i/3}) ]  \nonumber \\
& + x_{32} [ 2(1-e^{-2\pi i/3}) + (1-e^{2\pi i/3}) ] \} \delta z +3 x_{3} =0 .
\end{align}
To solve this one can either use the quadratic formula, or note that if we neglect the $\delta z$ term
this gives $\delta z \sim 1/\sqrt{N}$, and then making use of this we see that both the constant term
in the equation and $(\delta z)^{2}$ are of order $1/N$, while the term linear in $\delta z$ is of order
$1/N^{3/2}$, i.e. smaller than the other two, so that it represents an even smaller order correction.  This
implies that to lowest order we have
\begin{equation}
\delta z = \pm i \sqrt{3 x_{3}} .
\end{equation}
Define $\theta_{0}$ so that
\begin{equation}
e^{2i\theta_{0}}=1+i\sqrt{3x_{3}} ,
\end{equation}
which implies that $z=-e^{\pm 2i\theta_{0}}$ and that $2\theta_{0} = \sqrt{3x_{3}}$.  The eigenvalues
of $U$ corresponding to these values of $z$ are $\pm i e^{\pm i\theta_{0}}$.

We can now find the eigenstates corresponding to these four eigenvalues by making use of the
relations
\begin{equation}
\alpha_{j} = \frac{\eta}{\lambda^{2}e^{-i\phi_{j}}+1} \hspace {1cm} 
\beta_{j}=\lambda e^{-i\phi_{j}}\alpha_{j} ,
\end{equation}
where $\lambda$ is one of the eigenvalues, and $\eta$ is a normalization constant.  This gives 
us the four eigenvectors:
for $\lambda = ie^{i\theta_{0}}$
\begin{align}
|\xi_{1}\rangle = {} & \eta \left[ \sum_{j=1}^{n_{31}} \frac{|j,0\rangle + ie^{i\theta_{0}}
|0,j\rangle}{1-e^{2i\theta_{0}}} \right. \nonumber \\
& + \sum_{j=n_{31}+1}^{N/2} \frac{|j,0\rangle 
+ ie^{i\theta_{0}} e^{-2\pi i/3} |0,j\rangle}{1-e^{2i\theta_{0}}e^{-2\pi i /3}} \nonumber \\
& +  \sum_{j=(N/2)+1}^{(N/2)+n_{32}} \frac{|j,0\rangle 
 + ie^{i\theta_{0}}|0,j\rangle}{1-e^{2i\theta_{0}}} \nonumber \\
& \left. + \sum_{j=(N/2)+n_{32}+1}^{N} \frac{|j,0\rangle + ie^{i\theta_{0}} e^{2\pi i/3} |0,j\rangle}{1-e^{2i\theta_{0}} e^{2\pi i /3} } \right]  ,
\end{align}
for $\lambda = -ie^{i\theta_{0}}$ 
\begin{align}
|\xi_{2}\rangle = {} & \eta \left[ \sum_{j=1}^{n_{31}} \frac{|j,0\rangle - ie^{i\theta_{0}}
|0,j\rangle}{1-e^{2i\theta_{0}}} \right. \nonumber \\
& + \sum_{j=n_{31}+1}^{N/2} \frac{|j,0\rangle 
- ie^{i\theta_{0}} e^{-2\pi i/3} |0,j\rangle}{1-e^{2i\theta_{0}}e^{-2\pi i /3}} \nonumber \\
& +  \sum_{j=(N/2)+1}^{(N/2)+n_{32}} \frac{|j,0\rangle 
 - ie^{i\theta_{0}}|0,j\rangle}{1-e^{2i\theta_{0}}} \nonumber \\
& \left. + \sum_{j=(N/2)+n_{32}+1}^{N} \frac{|j,0\rangle - ie^{i\theta_{0}} e^{2\pi i/3} |0,j\rangle}{1-e^{2i\theta_{0}} e^{2\pi i /3} } \right]  ,
\end{align}
for $\lambda = ie^{-i\theta_{0}}$
\begin{align}
|\xi_{3}\rangle = {} & \eta \left[ \sum_{j=1}^{n_{31}} \frac{|j,0\rangle 
+ ie^{-i\theta_{0}} |0,j\rangle}{1-e^{-2i\theta_{0}}} \right. \nonumber \\
& + \sum_{j=n_{31}+1}^{N/2} \frac{|j,0\rangle 
+ ie^{-i\theta_{0}} e^{-2\pi i/3} |0,j\rangle}{1-e^{-2i\theta_{0}}e^{-2\pi i /3}} \nonumber \\
& +  \sum_{j=(N/2)+1}^{(N/2)+n_{32}} \frac{|j,0\rangle 
 + ie^{-i\theta_{0}}|0,j\rangle}{1-e^{-2i\theta_{0}}} \nonumber \\
& \left. + \sum_{j=(N/2)+n_{32}+1}^{N} \frac{|j,0\rangle + ie^{-i\theta_{0}} e^{2\pi i/3} |0,j\rangle}{1-e^{-2i\theta_{0}} e^{2\pi i /3} } \right]  ,
 \end{align}
and, finally, for $\lambda = -ie^{-i\theta_{0}}$
\begin{align}
|\xi_{4}\rangle = {} & \eta \left[ \sum_{j=1}^{n_{31}} \frac{|j,0\rangle 
- ie^{-i\theta_{0}} |0,j\rangle}{1-e^{-2i\theta_{0}}} \right. \nonumber \\
& + \sum_{j=n_{31}+1}^{N/2} \frac{|j,0\rangle 
- ie^{-i\theta_{0}} e^{-2\pi i/3} |0,j\rangle}{1-e^{-2i\theta_{0}}e^{-2\pi i /3}} \nonumber \\
& +  \sum_{j=(N/2)+1}^{(N/2)+n_{32}} \frac{|j,0\rangle 
 - ie^{-i\theta_{0}}|0,j\rangle}{1-e^{-2i\theta_{0}}} \nonumber \\
& \left. + \sum_{j=(N/2)+n_{32}+1}^{N} \frac{|j,0\rangle - ie^{-i\theta_{0}} e^{2\pi i/3} |0,j\rangle}{1-e^{-2i\theta_{0}} e^{2\pi i /3} } \right]  .
\end{align}
We find that for these vectors to be normalized, we must have $\eta = (1/2)\sqrt{3/N}$.
 
Now in order to help make sense of this, define the following four vectors
\begin{align}
|\zeta_{1}^{(\pm )}\rangle = {} & \frac{i}{\sqrt{2n_{3}}} \left[ \sum_{j=1}^{n_{31}} \left(|j,0\rangle 
\pm i|0,j\rangle \right) \right. \nonumber \\
& \left. + \sum_{j=(N/2)+1}^{(N/2)+n_{32}} \left(|j,0\rangle \pm i|0,j\rangle \right) \right] ,
\end{align} 
and 
\begin{align}
|\zeta_{2}^{(\pm )}\rangle = {} & \eta \sqrt{2} \left[ \sum_{j=n_{31}+1}^{N/2} \frac{|j,0\rangle \pm ie^{-2\pi i /3} |0,j\rangle}{1-e^{-2\pi i /3}} \right.  \nonumber \\
& \left. \sum_{j=(N/2)+n_{32}+1}^{N} \frac{|j,0\rangle 
\pm ie^{2\pi i /3}|0,j\rangle}{1- e^{2\pi i /3} } \right] .
\end{align}
Now we have, approximately, that
\begin{align}
|\xi_{1}\rangle = {} & \frac{1}{\sqrt{2}} \left(|\zeta_{1}^{(+)}\rangle + |\zeta_{2}^{(+)}\rangle \right) \nonumber \\
|\xi_{3}\rangle = {} & \frac{1}{\sqrt{2}} \left(-|\zeta_{1}^{(+)}\rangle + |\zeta_{2}^{(+)}\rangle \right) \nonumber \\
|\xi_{2}\rangle = {} &  \frac{1}{\sqrt{2}} \left(|\zeta_{1}^{(-)}\rangle + |\zeta_{2}^{(-)}\rangle \right) \nonumber \\
|\xi_{4}\rangle = {} &  \frac{1}{\sqrt{2}} \left(-|\zeta_{1}^{(-)}\rangle + |\zeta_{2}^{(-)}\rangle \right) .
\end{align}

Now let us consider for the initial state of the walk the state
\begin{equation}
|\psi_{init}\rangle = \frac{1}{\sqrt{N}}\sum_{j=1}^{N} |j,0\rangle .
\end{equation}
Defining the state
\begin{equation}
|\zeta_{0}\rangle = \frac{1}{\sqrt{2}}\left( |\zeta_{2}^{(+)}\rangle + |\zeta_{2}^{(-)}\rangle \right)
\end{equation}
we find that $|\langle \zeta_{0}|\psi_{init}\rangle |= \sqrt{3}/2$, i.e. these states have a substantial
overlap.  Noting that 
\begin{equation}
|\zeta_{0}\rangle = \frac{1}{2}\sum_{k=1}^{4} |\xi_{k}\rangle
\end{equation}
we see that after $m$ steps 
\begin{align}
|\zeta_{0}\rangle \rightarrow {} & \frac{1}{2} \left[ (ie^{i\theta_{0}})^{m}|\xi_{1}\rangle + (ie^{-i\theta_{0}})^{m}
|\xi_{3}\rangle \right. \nonumber \\
& \left. + (-ie^{i\theta_{0}})^{m}|\xi_{2}\rangle + (-ie^{-i\theta_{0}})^{m}|\xi_{4}\rangle \right]  \nonumber \\
= {} & \frac{1}{\sqrt{2}} \left\{ i^{m} \left[ i\sin (m\theta_{0}) |\zeta_{1}^{(+)}\rangle + \cos (m\theta_{0})
|\zeta_{2}^{(+)}\rangle  \right] \right.  \nonumber \\
& \left. + (-i)^{m} \left[ i\sin (m\theta_{0}) |\zeta_{1}^{(-)}\rangle + \cos (m\theta_{0}) |\zeta_{2}^{(-)}\rangle  \right] \right\} .
\end{align}
What this means is that when $m\theta_{0}=\pi /2$, $|\zeta_{0}\rangle$ becomes a state with the
particle located on the edges with $\phi = 0$.  Because the overlap between 
$U^{m}|\psi_{init}\rangle$ and $U^{m}|\zeta_{0}\rangle$ is the same as the overlap between 
$|\psi_{init}\rangle$ and $|\zeta_{0}\rangle$, if we measure the location of the particle in the state 
$U^{m}|\psi_{init}\rangle$, we will find that it is, with probability $3/4$ on one of the edges connected
to a vertex with $\phi = 0$.  Note that in order to find edges with $\phi = 0$ with this procedure, we 
do not have to know which edges have $\phi = 2\pi /3$ and which have $\phi = -2\pi /3$.  Also note
that unlike in the usual Grover algorithm, the probability to find a vertex with $\phi = 0$ does
not go to one as $N\rightarrow \infty$.

\section{Oracle version}
Now let us show that we can rephrase our walk search problem as an oracle search problem.
We consider the situation in which the function that the oracle evaluates is multivalued. The function $f(j)$ with $j=1, 2, \cdots, N$ can take on integer values between 0 and $(d-1)$, where $d$ is an integer greater than 2. It is 0 at $j=j_0$, while it takes on one of the integer values between 1 and $(d-1)$ at other values of $j$.  We are interested in finding $j_0$.  We will do this in two different ways.  The first reduces to a standard Grover search, 
and the probability of finding $j_{0}$ using this method is close to $1$. The second results in a search that is analogous to what happens in the quantum walk on the star graph, and in that case the probability of finding
$j_{0}$ is a decreasing function of $d$.

Let us first show how the oracle for a multivalued function can be used to perform a standard Grover search \cite{grover1, grover2}.  Our oracle has the action $O|j\rangle_{1} |k\rangle_{2} 
= |j\rangle_{1} |k\oplus f(j)\rangle_{2}$, where $|j\rangle_{1}$ is the state of the 
input register that contains the argument of the function, $|k\rangle_{2}$ is a qudit state,
and $\oplus$ denotes addition modulo $d$.  We need to manufacture an operation that flips the
sign of $|j_{0}\rangle$ but leaves other states $|j\rangle$, where $j\neq j_0$ alone.  The
operator $O^{-1}(I_{1}\otimes I_{2} -2I_{1}\otimes |0\rangle_{2}\langle 0|)O$, where $I_{m}$ for
$m=1,2$ is the identity on space $m$, does exactly that.  We have
\begin{align}
& O^{-1}(I_{1}\otimes I_{2} -2I_{1}\otimes |0\rangle_{2}\langle 0|)O |j\rangle_{1}|0\rangle_{2}
\nonumber \\
& =\left\{ \begin{array}{cc} |j\rangle_{1}|0\rangle_{2} & \hspace{5mm} j\neq j_{0} \\
-|j\rangle_{1}|0\rangle_{2} & \hspace{5mm} j=j_{0} \end{array} \right.   ,
\end{align}
because $j_{0}$ is the only value of $j$ for which $f(j)$ is $0$.  Note that because $O^{d}$ is the
identity, we have that $O^{-1}=O^{d-1}$ so that we can realize $O^{-1}$ simply by applying $O$
$d-1$ times.  If we now apply this operator followed by the usual inversion-about-the-mean
operator, we have the standard Grover iteration. This will result in a probability of close to one
of finding $j_{0}$.

That is not how our quantum walk worked, however.  There, we had $d$ different phase shifts and
different values of $f(j)$ led to different phase shifts.  Let us now formulate an oracle problem in
which that happens.

We assume for simplicity that $f(j)=0$ at only one particular value of $j$, say $j_0$.
To search for $j_0$, we adopt a straightforward extension of the standard Grover algorithm\cite{grover1, grover2}. The starting point of our search is to take a $d \times d$ matrix  
\begin{equation}
{H_{d}} =
\left( \begin{array}{ccccc}
1 & 1& 1& \cdots &  1 \\
1 & \beta & \beta^{2} & \cdots &  \beta^{d-1} \\
1 & \beta^{2} & \beta^{4} & \cdots &  \beta^{2(d-1)} \\
\vdots &  \vdots & \vdots & \ddots &  \vdots \\
1 & \beta^{d-1} & \beta^{2(d-1)} & \cdots &  \beta^{(d-1)^{2}}
\end{array} \right)
\end{equation}
as a $d$-dimensional equivalent of the Hadamard operator 
$H \equiv H_{2}=\left( \begin{array}{cc}
1 & 1 \\
1 & -1
\end{array} \right)$, where
\begin{equation}
\beta=e^{i2\pi /d} \;  (\beta^{d}=1),
\end{equation}
and
\begin{equation}
1+\beta +\beta^{2} + \cdots +\beta^{d-1} =0.
\end{equation} 
With $n$ qudits in the main register ($d^{n}=N$) and one ancilla qudit in the secondary register, the search is then conducted in the following order: 
\begin{enumerate}
\item Initial preparation. The $n$ qudits are prepared in state
\begin{equation} 
\left(H_{d} |0 \rangle \right)^{\otimes n}=\left(\frac{1}{\sqrt{d}} \sum_{k=0}^{d-1} |k \rangle \right)^{\otimes n} =\frac{1}{\sqrt{N}}\sum_{j=1}^{N} |j \rangle , 
\end{equation}
and the ancilla qudit in state $H_{d} |1 \rangle =\frac{1}{\sqrt{d}} \sum_{k=0}^{d-1} \beta^{k} |k \rangle$.       
\item Oracle transformation. The $n$ qudits along with the ancilla qudit are directed to an oracle in which the transformation $|j \rangle |k \rangle  \longrightarrow |j \rangle |f(j) \oplus k \rangle$ is performed , where $|j \rangle$ and $|k \rangle$ represent the state of the $n$ qudits in the main register and of the ancilla qudit in the secondary register, respectively. When the ancilla qudit is in state $\frac{1}{\sqrt{d}} \sum_{k=0}^{d-1} \beta^{k} |k \rangle$, the oracle transformation is represented by 
\begin{equation}
|j \rangle\left(\frac{1}{\sqrt{d}} \sum_{k=0}^{d-1} \beta^{k} |k \rangle \right)  \longrightarrow \beta^{-f(j)} |j \rangle \left(\frac{1}{\sqrt{d}} \sum_{k=0}^{d-1} \beta^{k} |k \rangle \right) . 
\end{equation}
The oracle transformation can thus be written, in the $N$-dimensional Hilbert space of the $n$-qudit state, as $O=\sum_{j=1}^{N} \beta^{-f(j)} |j\rangle \langle j|$.
\item Inversion about average. This is achieved by 
\begin{equation}
D=\frac{2}{N} \sum_{i=1}^{N} \sum_{j=1}^{N} |i \rangle \langle j| -\sum_{j=1}^{N} |j \rangle \langle j| .
\end{equation} 
\item Repeat the operation $G \equiv DO$ until the state of the $n$ qudits in the main register comes sufficiently close to the desired state $|j_0 \rangle $. 
\item Measure the state of the $n$ qudits in the main register to get the solution. Evaluate $f(j)$ to check that the solution is correct. If the solution is correct, then the search succeeds. If not, go back to (1) and start the search over again. This last step (5) is necessary because, as will be shown later, the probability for the search to produce the correct solution is less than 1 for $d \ge 3$. 
\end{enumerate}

In the $N$-dimensional Hilbert space, the $N \times N$ matrix $G=DO$ is given by
\begin{equation}
\label{groverop}
G=\frac{2}{N} \sum_{i=1}^{N} \sum_{j=1}^{N} \beta^{-f(j)} |i \rangle \langle j|-\sum_{j=1}^{N} \beta^{-f(j)} |j \rangle \langle j|.
\end{equation}
After applications of the operator $G$ $k$ times, the state of the $n$ qudits in the main register becomes 
$|\psi_{k} \rangle =G^{k} |\psi_{0} \rangle$, 
where 
$|\psi_{0} \rangle =\frac{1}{\sqrt{N}} \sum_{j=1}^{N} |j \rangle$.
The probability to find the desired state $|j_0 \rangle$ after $k$ iterations is then given by
\begin{equation}
\label{succ-prob}
P_{k} = \left|\langle j_0 |G^{k} |\psi_0 \rangle \right|^{2} = \left|\sum_{j=1}^{N} z_{j}^{k} \langle j_0 |\chi_j \rangle \langle \chi_j |\psi_0 \rangle \right|^{2} ,
\end{equation}
where $z_j $'s and $|\chi_j \rangle $'s are eigenvalues and eigenvectors of the matrix $G$. The probability $P_k $  can thus be determined by calculating $z_{j}$'s and $|\chi_{j} \rangle$'s. When $d=2$, the method described here reproduces the results of the standard Grover search with a two-valued function. 

Let us now find the eigenvalues $z_{k}$ ($k=1, 2, \cdots, N$) and corresponding eigenvectors 
$|\chi_k \rangle $ of the $N \times N$ matrix $G$ given by Eq. (\ref{groverop}), where
\begin{equation}
\label{G-eigen}
G|\chi_k \rangle =z_{k} |\chi_{k} \rangle, \; k=1, 2, \cdots, N
\end{equation}
The eigenvectors $|\chi_{k} \rangle$ can be expressed in a linear superposition of the basis vectors $|j \rangle$ as
\begin{equation}
\label{gen-eigen1}
|\chi_{k} \rangle =\sum_{j=1}^{N} c_{kj} |j \rangle
\end{equation}
Substituting Eqs. (\ref{groverop}) and (\ref{gen-eigen1}) into Eq. (\ref{G-eigen}) and solving the resulting equation for $c_{kj}$, we obtain
\begin{equation}
\label{gen-eigen2}
c_{kj} = \frac{\frac{2}{N} S_{k}}{\beta^{-f(j)} +z_{k}}, \quad j=1, 2, \cdots, N
\end{equation} 
where
\begin{equation}
S_{k} =\sum_{j=1}^{N}  \beta^{-f(j)} c_{kj}
\end{equation}
Substituting our expression for $c_{jk}$ back into the above equation, we obtain
\begin{equation}
\label{eigsum}
\sum_{j=1}^{N}  \frac{1}{1+\beta^{f(j)} z_{k} } =\frac{N}{2}
\end{equation}
This is identical to Eq. (6). We note that one application of the Grover operator $G$ is equivalent to two steps of the walk, and therefore the eigenvalue $z$ of $G$ is equal to the square of the eigenvalue, $\lambda^2$, of the operator $U$. We proceed to solve Eq. (\ref{eigsum}) for the case where there is one $j$, i.e., $j_0$, at which $f(j)=0$. We further assume for simplicity that the integer values between 1 and $(d-1)$ of the function $f(j)$ are distributed evenly among other $(N-1)$ $j$'s. There is thus one $j$, i.e., $j_0$, for which $f(j)=0$ and $\frac{N-1}{d-1}$ $j$'s each at which $f(j)=1, 2, \cdots ,d-1$.  We assume that $\frac{N-1}{d-1}$ is an integer. Note that if we choose $N=d^{n}$, where $n$ is a positive integer, then $\frac{N-1}{d-1}$ is an integer. 
Using the relation
\begin{equation}
\label{geomser}
1+x+x^2 +\cdots +x^{d-1} =\frac{1-x^{d}}{1-x}
\end{equation}
with $x=-\beta^{j} z_{k}$ and noting that $\beta^{jd}=1$, Eq. (\ref{eigsum}) can be rewritten as 
\begin{equation}
\frac{1}{1+z_{k}} +\left(\frac{N-1}{d-1}\right)\frac{1}{1-(-z_{k})^{d}}\left[d-1-\sum_{j=1}^{d-1} (-z_{k})^{j} \right]=\frac{N}{2} ,
\end{equation}
or
\begin{equation}
\label{qudit-fin-N}
(-z_{k})^{d} -\left(\frac{2}{d-1}\right)\left(\frac{N-d}{N}\right) \sum_{j=1}^{d-1}(-z_{k})^{j} +1 =0 .
\end{equation}

In the limit $N \rightarrow \infty$, the above equation becomes 
\begin{equation}
\label{Ninf}
(-z_{k})^{d} -\left(\frac{2}{d-1}\right) \sum_{j=1}^{d-1}(-z_{k})^{j} +1 =0 .
\end{equation}
This equation, because it has real coefficients, will have as its roots either real numbers or complex
conjugate pairs.  By examining the above expression and its derivative, both of which must be
zero at a double root, one finds that the only double root of this equation is $z_{k}=-1$.  This can be
seen as follows.  Setting the derivative of the above equation equal to zero and then multiplying by
$-z_{k}$ gives
\begin{equation}
(-z_{k})^{d} -\frac{2}{d(d-1)}\sum_{j=1}^{d-1} j(-z_{k})^{j} =0 .
\end{equation}
Substituting into this equation for $(-z_{k})^{d}$ from Eq. (\ref{Ninf}), we find
\begin{equation}
\label{deriv-cond}
\frac{2}{d-1} \sum_{j=1}^{d-1}\left( 1-\frac{j}{d}\right) (-z_{k})^{j} =1 .
\end{equation}
Noting that 
\begin{equation}
\frac{2}{d-1} \sum_{j=1}^{d-1}\left( 1-\frac{j}{d}\right) =1 ,
\end{equation}
we see that the only way Eq. (\ref{deriv-cond}) can be satisfied is if all of the terms add in phase.  The
only way that can happen is if $z_{k}=-1$, and so, as stated above, that is the only double root.

Going back
to the equation for finite $N$, we find that this root splits into two roots.  In order to obtain an expression for these two roots, we substitute $z_{k}=-1+\delta z$ into Eq. (\ref{qudit-fin-N}) and find $\delta z  \approx \pm i\sqrt{\frac{12}{N(d+1)}}$. We designate these two roots as $z_{1}$ and $z_{2}$. We can now write
\begin{eqnarray}
z_{1} \approx -1+i\sqrt{\frac{12}{N(d+1)}} \equiv -e^{-i2\theta} \\
z_{2} \approx -1-i\sqrt{\frac{12}{N(d+1)}} \equiv -e^{i2\theta}
\end{eqnarray}
where
\begin{equation}
\sin (2\theta) \approx 2\theta \approx \sqrt{\frac{12}{N(d+1)}}.
\end{equation}
The finite $N$ corrections of these roots are of order $N^{-1/2}$, whereas the finite $N$ corrections to
the other roots of Eq. (\ref{Ninf}) are of order $N^{-1}$.

Let us now find the eigenvectors $|\chi_{1} \rangle$ and $|\chi_{2} \rangle$ corresponding to $z_{1}$ and $z_{2}$. Using Eqs. (\ref{gen-eigen1}) and (\ref{gen-eigen2}), we write
\begin{equation}
|\chi_{1} \rangle =\eta_1 \sum_{j=1}^{N}  \frac{1}{\beta^{-f(j)} -e^{-i2\theta}}|j \rangle
\end{equation}
and similarly for $|\chi_{2} \rangle$. The normalization condition $1=\langle \chi_{1} |\chi_{1} \rangle $ determines the constant $\eta_{1}$. Using again Eq. (\ref{geomser}) with $x=\beta^{-j} e^{i2\theta}$, the normalization condition is reduced to
\begin{widetext}
\begin{equation}
1 = |\eta_{1}|^{2} \left\{ \frac{1}{2(1-\cos 2\theta)} + \left(\frac{N-1}{d-1}\right)\frac{1}{2[1-\cos (2d\theta )]}[d(d-1)-2\sum_{j=1}^{d-1} (d-j) \cos (2j\theta )] \right\}
\end{equation}
\end{widetext}
For $(n/N)\ll 1$, we have that $\cos 2n\theta \approx 1-n^{2} \frac{6}{N(d+1)}$, we obtain $\eta_{1} \approx \sqrt{\frac{6}{N(d+1)}}$, and thus
\begin{equation}
\label{chi1}
|\chi_{1} \rangle \approx \sqrt{\frac{6}{N(d+1)}}\sum_{j=1}^{N}  \frac{1}{\beta^{-f(j)} -e^{-i2\theta}}|j \rangle
\end{equation}
and similarly
\begin{equation}
|\chi_{2} \rangle \approx \sqrt{\frac{6}{N(d+1)}}\sum_{j=1}^{N}  \frac{1}{\beta^{-f(j)} -e^{i2\theta}}|j \rangle
\end{equation}
Other eigenstates are given by
\begin{equation}
|\chi_{k} \rangle =\eta_k \sum_{j=1}^{N}  \frac{1}{\beta^{-f(j)} +z_{k}}|j \rangle, \quad k=3, 4, \cdots, N
\end{equation}
and the normalizations for these states are
\begin{equation}
\left|\eta_k\right|^2 \sum_{j=1}^{N}  \left| \frac{1}{\beta^{-f(j)} +z_{k}} \right|^2 =1.
\end{equation}
For any $j$ and $k$, we have $\left|\beta^{-f(j)} +z_{k}\right|<2$. Therefore, $\eta_k$'s should satisfy
\begin{equation}
\left|\eta_k\right|^2 \frac{N}{4} < 1
\end{equation}
or
\begin{equation}
\eta_k < \frac{2}{\sqrt{N}}.
\end{equation}

The probability $P_{k}$ to find the desired state $|j_{0} \rangle$ after $k$ iterations is given by Eq. 
(\ref{succ-prob}).  In order to evaluate this, we begin by noting that
\begin{equation}
\langle j_0 |\chi_1 \rangle \approx -i \frac{1}{\sqrt{2}}, \quad \langle j_0 |\chi_2 \rangle \approx i \frac{1}{\sqrt{2}} ,
\end{equation}
which implies that $\langle j_{0}|\chi_{k}\rangle \approx 0$.  Therefore, we have that
\begin{equation}
P_{k} \approx \left|z_{1}^{k} \langle j_0 |\chi_{1} \rangle \langle \chi_1 |\psi_0 \rangle+z_{2}^{k} \langle j_0 |\chi_{2} \rangle \langle \chi_2 |\psi_0 \rangle\right|^{2}
\end{equation}
From Eq. (\ref{chi1}), we have
\begin{equation}
\langle \chi_1 |\psi_0 \rangle \approx \sqrt{\frac{6}{N^2 (d+1)}} \sum_{j=1}^{N}  \frac{1}{\beta^{f(j)} -e^{i2\theta}}
\end{equation}
and similarly for $\langle \chi_2 |\psi_0 \rangle$. Evaluating the summation in in this equation using Eq. (\ref{eigsum}), we obtain
\begin{equation}
\langle \chi_1 |\psi_0 \rangle =\langle \chi_2 |\psi_0 \rangle \approx -\sqrt{\frac{3}{2(d+1)}}
\end{equation}
Making use of these results, we finally obtain for $P_{k}$
\begin{equation}
P_{k} \approx \frac{3}{d+1} \sin^{2} (2k\theta) .
\end{equation}
The maximum probability is given by $P_{max} \approx \frac{3}{d+1}$ and the number of iterations to reach the maximum
 probability is $k_{max} \approx \frac{\pi}{4\theta} \approx \frac{\pi}{4} \sqrt{\frac{N(d+1)}{3}}$. Since the maximum probability decreases as $d$ increases, it gets increasingly important to check in the end that the solution produced by the search is indeed correct. This can be done either classically or quantum mechanically. When $d=3$, we have $P_{max} \approx \frac{3}{4}$, in agreement with the result presented in the previous section. Although the maximum probability decreases with $d$, the number  $k_{max}$ of iterations required to reach the maximum probability still scales as $\sqrt{N}$ for $d \ge 3$. This indicates that the quantum walk search still provides quadratic speedup, the best one can gain from a quantum search algorithm \cite{bennett, boyer, zalka}, for $d \ge 3$. 

So far, we have assumed that there is a single match. We now briefly comment on the situation where there are multiple matches \cite{boyer}. If the number $M$ of matches is small compared with the total number $N$, only a trivial modification is needed. Since the algebra involved is straightforward, we here state the result only. The maximum probability $P_{max}$ remains the same as long as $M \ll N$, but the number $k_{max}$ of steps to reach the maximum probability decreases with $M$. In fact, we obtain $P_{k} \approx \frac{3}{d+1} \sin^{2} (2k\theta)$, where $\sin\theta=\sqrt{\frac{3M}{N(d+1)}}$, and therefore $k_{max} \approx \frac{\pi}{4} \sqrt{\frac{N(d+1)}{3M}}$. The number $k_{max}$ scales as $1/\sqrt{M}$. Classically, when there are $M$ matches, one would expect to find a match $M$ times faster than when there is a single match. One thus expects that classically the number $k_{max}$ to scale as $1/M$. It is then natural to expect that this number would scale as $1/\sqrt{M}$ for the quantum search which is quadratically faster than any classical search. If the number $M$ of matches is not small compared with $N$, a different search strategy from that described so far should be employed because a match can be found with a reasonably high probability after only a small number of iterations. In general, the state of the qudits in the main register changes rapidly, and it seems difficult to find a general rule here. One simply needs to find the best strategy for a given situation. 

\section{Conclusion}\label{conclusion}

We have studied a quantum walk search on a star graph and an analogous oracle search problem.  The quantum
walk is trying to find the set of vertices that reflect the particle with a particular phase, and the oracle search is
trying to find the input values for a multivalued function that produce a particular output value.
We have shown that the maximum probability $P_{max}$ for the search on a star graph to find the correct vertex decreases with $d$, the number of different phases with which the particle is reflected from outer vertices.  We have also shown that the number $k_{max}$ of steps needed to reach the maximum probability increases with $d$. These two effects combine to increase the average number of steps required to be taken in order to find the correct vertex. Despite that, however, this number still scales as $\sqrt{N}$ and the quantum walk search still gives the quadratic speedup over a classical search for any value of $d$ as long as $N \gg d$.

We examined two different oracle searches.  One mimicked the quantum walk, and for this search the probability of finding the inputs that produced a particular output decreased with $d$, just as for the quantum walk.  However, we also showed that the oracle can be used in a different way so that the  probability of finding the desired elements is close to one.  This indicates that the use of the oracle is more powerful than the quantum walk search we studied.

Finally, we would like to note that it has been possible in some cases to improve the maximum probability of finding a marked element in a quantum walks search.  The original quantum-walk search on the hypercube due to Shenvi, Kempe and Whaley found the marked vertex with a probability of less than $1/2$ \cite{shenvi}.  Recently, methods have been found to improve this probability \cite{potocek}.  Whether similar methods could be used to improve the search on the star graph is a subject for further research.

\section*{Acknowledgments}

This research has been supported by the SBS Foundation of Korea and by the National Science Foundation under grant number PHY-0903660.

\end{document}